\documentclass[12pt]{article}
\usepackage{graphicx}
\usepackage[english]{babel}
\usepackage{hyperref}
\usepackage{xcolor}
\begin{document}
\pagestyle{empty}

\begin{center}
{\Huge\bf 
Gamma Factory at CERN 
\vspace{2mm}\\
-- novel research tools made of light$^{\star}$
}
\end{center}
\vskip 3mm
\begin{center}
W.~P{\l}aczek$^{1}$,
A.~Abramov$^{2}$,
S.E.~Alden$^{2}$,
R.~Alemany Fernandez$^{3}$,
P.S.~Antsiferov$^{4}$, 
A.~Apyan$^{5}$, 
H.~Bartosik$^{3}$,
E.G.~Bessonov$^{6}$,
N.~Biancacci$^{3}$,
J.~Biero\'n$^{1}$,
A.~Bogacz$^{7}$,
A.~Bosco$^{2}$,
R.~Bruce$^{3}$,
D.~Budker$^{8,9,10}$,
K.~Cassou$^{11}$,
F.~Castelli$^{12}$, 
I.~Chaikovska$^{11}$,
C.~Curatolo$^{13}$,
P.~Czodrowski$^{3}$,
A.~Derevianko$^{14}$,
K.~Dupraz$^{11}$,
Y.~Dutheil$^{3}$, 
K.~Dzier\.z\c{e}ga$^{1}$,
V.~Fedosseev$^{3}$,
N.~Fuster Martinez$^{3}$, 
S.M.~Gibson$^{2}$,
B.~Goddard$^{3}$, 
A.~Gorzawski$^{15,3}$,
S.~Hirlander$^{3}$, 
J.~Jowett$^{3}$, 
R.~Kersevan$^{3}$, 
M.~Kowalska$^{3}$,
M.W.~Krasny$^{16,3}$,
F.~Kroeger$^{17}$,  
M.~Lamont$^{3}$, 
T.~Lefevre$^{3}$,
D.~Manglunki$^{3}$,
B.~Marsh$^{3}$,
A.~Martens$^{11}$, 
J.~Molson$^{3}$,
D.~Nutarelli$^{11}$,
L.J.~Nevay$^{2}$,
A.~Petrenko$^{3}$,
V.~Petrillo$^{12}$,
S.~Radaelli$^{3}$, 
S.~Pustelny$^{1}$,
S.~Rochester$^{18}$,
M.~Sapinski$^{19}$,
M.~Schaumann$^{3}$,
L.~Serafini$^{12}$,
V.P.~Shevelko$^{6}$, 
T.~Stoehlker$^{17}$, 
A.~Surzhikov$^{20}$
I.~Tolstikhina$^{6}$, 
F.~Velotti$^{3}$,
G.~Weber$^{17}$, 
Y.K.~Wu$^{21}$,
C.~Yin-Vallgren$^{3}$,
M.~Zanetti$^{22,13}$,
F.~Zimmermann$^{3}$, 
M.S.~Zolotorev$^{23}$ 
\\and 
F.~Zomer$^{11}$  

\vskip 3mm
{\em
${}^{1}$ Marian Smoluchowski Institute of Physics, Jagiellonian University, Krak\'ow, Poland \\
${}^{2}$ Royal Holloway University of London Egham, Surrey, TW20 0EX, UK\\
${}^{3}$ CERN, Geneva, Switzerland \\
${}^{4}$ Institute of Spectroscopy, Russian Academy of Sciences, Troitsk, Russia\\
${}^{5}$ A.I.~Alikhanyan National Science Laboratory,Yerevan, Armenia\\
${}^{6}$ P.N.~Lebedev Physical Institute, Russian Academy of Sciences, Moscow, Russia\\
${}^{7}$ Center for Advanced Studies of Accelerators,  Jefferson Lab, USA\\
${}^{8}$ Helmholtz Institute,  Johannes Gutenberg University, Mainz, Germany\\
${}^{9}$ Department of Physics, University of California, Berkeley, USA\\
${}^{10}$ Nuclear Science Division, E.O.~Lawrence National Laboratory, Berkeley, USA\\
${}^{11}$ LAL, Univ.\ Paris-Sud, CNRS/IN2P3, Universit\'e Paris-Saclay, Orsay, France\\
${}^{12}$ Department of Physics, INFN--Milan  and University of Milan,  Milan, Italy\\
${}^{13}$ INFN--Padua,  Padua, Italy\\
${}^{14}$ University of Nevada, Reno, Nevada 89557, USA\\
${}^{15}$ University of Malta, Malta\\
${}^{16}$ LPNHE, University Paris Sorbonne, CNRS--IN2P3, Paris, France\\
${}^{17}$ HI Jena, IOQ FSU  Jena and GSI Darmstadt, Germany\\
${}^{18}$ Rochester Scientific, LLC, El Cerrito, CA 94530, USA\\
${}^{19}$ GSI, Helmholtzzentrum f\"ur Schwerionenforschung, 64291 Darmstadt, Germany\\
${}^{20}$ Braunschweig University of Technology and Physikalisch-Technische Bundesanstalt, Germany \\
${}^{21}$ FEL Laboratory, Duke University,  Durham, USA\\
${}^{22}$ University of Padua, Padua, Italy\\
${}^{23}$ Center for Beam Physics, LBNL, Berkeley, USA
}
\\
\end{center}

\vspace{2mm}
\footnoterule
\noindent
{\footnotesize
$^{\star}$ Presented by W.~P{\l}aczek at the XXV Cracow Epiphany Conference on Advances in Heavy Ion Physics, 8--11 January 2019, Cracow, Poland. 
}


\newpage
\pagestyle{plain}
\setcounter{page}{1}

\begin{abstract}
\noindent
We discuss the possibility of creating novel research tools by producing
and storing highly relativistic beams of highly ionised atoms in the CERN accelerator complex, 
and by exciting their atomic degrees of freedom with lasers to produce high-energy photon beams.
Intensity of such photon beams would be by several orders of magnitude higher than offered by the presently
operating light sources, in the particularly interesting $\gamma$-ray energy domain of $0.1$--$400\,$MeV. 
In this energy range, the high-intensity photon beams can be used to produce secondary beams 
of polarised electrons, polarised positrons, polarised muons, neutrinos, neutrons and radioactive ions. 
New research opportunities in a wide domain of fundamental and applied physics can be opened 
by the Gamma Factory scientific programme based on the above primary and secondary beams.
\end{abstract}



\section{Introduction}

The main goal of the Gamma Factory (GF)  initiative  \cite{Krasny:2015ffb,Krasny:2018alc} is to create and store at CERN's accelerator complex 
partially-stripped ion (PSI) beams and to exploit their atomic degrees of freedom. 
Such beams can be accelerated at the SPS and the LHC to very high energies over a wide range of the relativistic Lorentz factor, $ 30 < \gamma_L < 3000$, 
and can reach high bunch intensities, $ 10^8 < N_{\rm bunch} <10^9$, as well as high bunch repetition rates, up to $20$~MHz.  
They can be excited by lasers tuned to atomic transition frequencies of the corresponding partially-stripped ions to produce highly energetic and highly collimated
$\gamma$-ray beams as a result of subsequent spontaneous photon emissions.   
Resonant excitations of atomic levels are possible even for heavy ions, like ${\rm Pb}$, owing to their ultra-relativistic velocities, which provide a Doppler-effect boost 
of laser-photon frequency  in the ion rest frame by a factor up to $2 \gamma _L$. 
Spontaneously emitted photons moving in the direction of the PSI beam have their LAB-frame energy boosted by a further factor of $2 \gamma _L$, 
so that the full process of absorption and emission results in a frequency amplification of the initial laser photon by a factor of up to $4 \gamma _L^2$. 
For LHC, this opens the possibility of producing $\gamma$-rays with energies up to $\sim 400\,$MeV.

The resonant absorption cross section can reach the gigabarn level, to be compared to the barn-level cross section for the inverse Compton scattering 
used in conventional $\gamma$-ray sources. Therefore, the intensity of $\gamma$-rays to be achieved in GF is by many orders of magnitude higher than
in the existing facilities and comparable to that of $X$-rays in the DESY XFEL. 

Photon absorption and spontaneous emission by partially stripped ions opens new possibilities of  beam cooling 
\cite{Bessonov:1995hd,ZolotorevBudker:1999,Schramm:2005fx}. 
Contrary to proton bunches, charged atomic beams can be efficiently cooled using Doppler-cooling techniques mastered over the last three decades by 
the atomic physics community, allowing to compress their bunch sizes and energy spread.

The high-energy and high-intensity $\gamma$-ray beam can be scattered on stationary targets to produce, with high efficiency,
secondary beams of polarised electrons, polarised positrons, polarised muons, neutrinos, neutrons and radioactive ions. 
These beams can offer a variety of new research opportunities at CERN in a wide area of fundamental as well as applied physics.

A detailed objective of the GF initiative is development of techniques and beams for use in specific applications. 
Given the available range of ion species, synchrotron energies, laser frequencies, interaction processes and secondary beams, 
this application range is potentially very wide. The identified domains are presented in Sections~2--4, 
with the emphasis on the gains offered by the GF approach. 
Three broad categories are considered: atomic beams, photon beams and secondary beams.
In Section~5, we present the GF project milestones. Tests of production, acceleration and storage of the atomic beams carried out in the SPS and the LHC 
in the years 2017--2018 are described in Section~6. In Section~7 we discuss development of software tools dedicated to GF. Goals of the planned GF  proof-of-principle 
experiment in the SPS are presented in Section~8. Finally, Section~9 contains a brief summary.

\section{Atomic beams}

\subsection{Beams for atomic, molecular and optical physics research} 

High-energy beams of highly charged high-$Z$ atoms, such as hydrogen-, helium- or lithium-like lead (${\rm Pb}^{81+}$, ${\rm Pb}^{80+}$, ${\rm Pb}^{79+}$),
are of particular interest for the Atomic, Molecular and Optical (AMO) physics community but have, so far, 
never been technologically attainable as research tools  \cite{Safronova:2017xyt}. 
The principal merits of such beams are:
\begin{itemize} 
 \item 
 strong electric field binding the electrons to the nucleus (exceeds that of the hydrogen atom by up to five orders
 of magnitude) providing unprecedented sensitivity to QED-vacuum effects, e.g.\ the Lamb shift;
\item 
 high amplification of the weak-interaction effects (up to nine orders of magnitude w.r.t. the hydrogen atom)  allowing 
 to study atomic and nuclear weak  interactions with unprecedented precision;
\item
 straightforward interpretation of experiments with hydrogen-like or helium-like atoms --  
 with the simplicity and precision of theoretical calculations inaccessible for multi-electron atoms,  
 \item 
 high-energy atomic transitions of highly charged ions can be excited by laser photons owing to  
 the Doppler effect -- the  large $\gamma_L$-factor of the beams compensates for the $\sim Z^2$ 
 increase of the binding energies,
\item 
 residual (or injected) gas molecules in the storage ring can be used to excite the atoms, allowing  
 for  precise studies of their emission spectra. 
\end{itemize}

The AMO research highlights include:
(1) studies of the basic laws of physics,
e.g.\ Lorenz invariance, the Pauli exclusion principle, CPT symmetries,
(2) precise measurements of the weak-interaction mixing angle
in the low-momentum-transfer regime, 
(3) measurements  of the nuclear charge radius and neutron skin depths in high-$Z$ nuclei 
and (4) searches for dark matter particles using 
the AMO detection techniques -- complementary to those used in particle physics.  

\subsection{Isoscalar ion beams for precision electroweak physics at LHC} 

Isoscalar nuclei, such as ${\rm Ca}$ or ${\rm O}$, i.e.\ containing the same number of protons and neutrons,
are  optimal for the LHC electroweak (EW) precision measurement programme. 
The relationship between the $W^+$, $W^-$  and $Z$ bosons production spectra for 
isoscalar beams simplifies the use of the $Z$-boson as a precision ``standard candle'' for 
the $W$-boson production processes,
e.g.\ it would allow to measure the $W$-boson mass at the LHC with the precision 
better than $10\,$MeV/c$^2$ \cite{Krasny:2010vd}. 
Unfortunately, this advantage is diminished by much lower luminosities that can
be achieved for such beams with respect to the proton ones using the present production schemes.

With GF, however, one can increase  the nucleon--nucleon collision luminosity by applying the Doppler-cooling
method to partially-stripped isoscalar ion beams in the SPS, followed by the stripping of the remaining electrons in the 
transfer line between the SPS and the LHC. 
If the transverse beam emittance of the colliding LHC beams of ${\rm Ca}$ or ${\rm O}$ ions is reduced by a factor of $10$, 
then the effective nucleon--nucleon luminosity in the respective collision modes could approach 
the nominal $pp$-collision luminosity.

\subsection{Electron beam for $ep$ operation of LHC}

The hydrogen-like or helium-like lead beams can be used as the 
carriers of the effective electron beams at the LHC. 
Colliding of such PSI beams with the counter-propagating proton beam would
produce both the proton--lead and the electron--proton collisions. 
The LHC could thus become an effective parasitic electron--proton(ion) collider (PIE) \cite{Krasny:2004ue}.  
The PIE $ep$ collider could reach the centre-of-mass energy of $200\,$GeV and the luminosity of
$10^{29}$~cm$^{-2}$s$^{-1}$, both inferior to the ones of the HERA
collider, but sufficiently large  for a precise, in situ, detector-dependent 
diagnostics of the partonic-beam emittance in the low Bjorken-$x$ region 
and a percent-level calibration of the luminosity of the proton--nucleus collisions at the LHC.  

\subsection{Driver beams for plasma wakefield acceleration}

High-intensity PSI beams could be efficient driver beams  
for hadron-beam-driven plasma wakefield acceleration \cite{Caldwell:2008ak}. 
Applying the Doppler beam-cooling technique would help to reduce the  emittance of the driver PSI beam, allowing 
to increase an acceleration rate of the witness beam. Alternatively, 
temporal modulation of the laser pulse could imprint the driver beam with the 
required micro-bunch structure.
In addition, the PSI beams would carry the ``ready-to-accelerate" electrons.
These electrons, exploited initially in the cooling process of the driver beam, 
could subsequently be used -- after stripping -- to form a precisely synchronised witness bunch. 

Our preliminary studies show that the reduction of the beam emittance by at least 
an order of magnitude and the cooling time below tens of seconds
are feasible both for the SPS and LHC PSI beams. 

\section{Photon beams}

In GF, lasers are used to excite resonant atomic transitions of relativistic partially-stripped ions,
resulting in spontaneously emitted photons. In this way, in the LAB frame one can achieve the frequency boost 
of the initial photons up to a factor of $4 \gamma _L^2$, which corresponds to producing the
outgoing  photon beam in a broad energy range up to $400\,$MeV.
The resonant photon-absorption cross section by PSI is higher by a factor up to $10^9$ than 
the inverse-Compton photon scattering cross section and, as a consequence, the PSI-beam-driven $\gamma$-ray source
intensity can be higher than that of electron-beam-driven ones by many orders of magnitude. 

\subsection{Gamma ray source}

The currently operating high-intensity Free Electron Laser (FEL) light sources 
produce photon beams with energies up to  $\approx 25\,$keV. The GF
photon beams could extend the energy reach by  four orders 
of magnitude with the FEL-like beam intensities. 
While the FEL-photon beams are optimal 
to study the atomic and molecular structure of matter, the GF beams would
allow to resolve the structure of  atomic nuclei. In addition, they 
have sufficient energies and intensities to produce secondary beams of matter particles. 
 
A very high $\gamma$-ray flux, corresponding to up to $10^{24}$ photons per year, 
would sizeably increase the sensitivity of searches for very weakly interacting dark matter particles 
in the ``anthropological'' keV--MeV mass region, e.g. in a beam-dump light-shining-through-the-wall 
type of experiment with the use a broad-energy-band photon beam.
 
\subsection{Photon collision schemes} 

High-intensity and high-brilliance photon beams produced by GF
can be used to realise, for the first time, two types of photon--photon colliders at CERN:
\begin{itemize}
\item 
an {\it elastic photon scattering} collider (below the $e^{\pm}$-pair production threshold), covering the centre-of-mass energy range
up to $\sim 100\,$keV, to be achieved by colliding the GF photon beam with the laser photons stacked in a Fabry--P\'erot cavity,
\item
 a {\it matter-particle producing} collider, covering the energy range up to $\sim 800\,$MeV,  
 to be achieved by colliding of the GF photon beam with its counter-propagating twin photon beam.  
\end{itemize}
With the former collider scheme one can explore the domain of fundamental QED measurements, such as  the elastic
large-angle light-by-light scattering produced with a rate of up to $\sim 1000$ events per second 
(to be compared with only tens of such events detected per year of the present LHC operation). 
The latter collider scheme, with the centre-of-mass energy sufficient to produce not only the opposite-charge lepton-pairs ($e^+e^-$ or $\mu^+\mu^-$) 
but also the $u$, $d$ and $s$ quark--antiquark pairs, can be used to explore colour-confinement phenomena at the colour-production threshold. 

In the case of detection of an axion-like particle or dark-photon signal in the shining-through-the-wall 
experiment, one of the above two collider schemes could be used a ``Dark Matter Production Factory'' by 
maximising the resonant production rate of dark-matter particles through a suitable choice of the photon-beam energy. 

\section{Gamma-ray-driven secondary beams} 

The GF photon beam can be extracted from its production zone and collided with external targets 
to produce secondary beams of matter particles. Such a method represents a change 
of a paradigm for the secondary beams production: from  the {\it mining} scheme, 
in which the dominant fraction of the primary-beam energy is wasted,     
to the {\it precision} or {\it production-by-demand} scheme.  

The  GF secondary-beam production scheme is based on peripheral 
(with small transfer-momentum) electromagnetic collisions of the photon beam with atoms of the target material.
As a consequence, a large fraction of the wall-plug power 
delivered to the PSI-beam storage ring
for continuous production of the primary photon beam could be 
transmitted to a given type of the secondary beam.  
Such a method could reduce considerably the target-heat load at a fixed 
intensity of the secondary beam, facilitating 
its design and overcoming the major technological challenges which limit the 
intensities of the proton-beam-driven muon, neutrino and neutron beams.  

\subsection{Polarised electron, positron and muon beams} 

The high-intensity $\gamma$-ray beam can be converted into 
high-intensity beams of $e^+$ and $e^-$. If the photon energy 
exceeds the muon-pair production threshold, the secondary beam will also contain an
admixture of $\mu^+$ and $\mu^-$.  
Beams of different lepton flavours can be easily separated with the time-of-flight method, since the produced
$e^{\pm}$ will be ultra-relativistic while $\mu^{\pm}$ will be non-relativistic (created close to the threshold).

In the process of the resonant absorption and the spontaneous emission by an atom with a spin-$0$ nucleus 
in which the electrons-spin state remains unchanged, the laser-photon polarisation is transferred to the emitted photon.
If the circularly polarised photon is converted into the lepton-pair in the electromagnetic field of such an atom (or nucleus), 
it will produce longitudinally polarised leptons.  

With the present CERN accelerator infrastructure and currently available laser technology,
the intensity of the GF source of the polarised electrons/positrons could reach $10^{17}$ positrons 
per second, which is at least three orders of magnitude higher than for the existing positron 
sources.  This would largely satisfy the source requirements for the ILC and CLIC colliders, or for the future 
high-luminosity  $ep$ ($eA$) collider project based on an energy-recovery linac. 

The target intensity of the GF polarised-muon beam is $10^{12}$ muons per second,
which is four orders of magnitude higher than in the presently best muon-beam facility. 
Two schemes are possible for the GF muon-beam production. In the first one, 
the $\gamma$-ray beam is tuned to its top energy range and $\mu^{\pm}$-pairs are produced in
the photon-conversion process on the stationary-target atoms. Without high-energy LHC, this would require a significant leap in the
intensity and bandwidth of a dedicated FEL source of $\sim 100\,$nm wavelength photons and an upgrade 
of the circumferential voltage of the LHC RF system.  In the second scheme,  first the positron bunches 
are produced by a low-energy photon beam and then they are accelerated in the dedicated positron ring 
to the energy exceeding the muon-pair production threshold in their collisions with a stationary target, 
i.e.\ $E_e \sim  2 m_{\mu}^2/m_e$ \cite{Antonelli:2015nla}.
The advantage of this scheme is that no LHC upgrade is  necessary and the conventional laser technology is sufficient. 
The intensity of the muon beam in these two schemes would be inferior to the proton-beam-driven muon sources. 
However, the product of the beam longitudinal and transverse emittances would be smaller by at least four orders 
of magnitude than that for the pion-decay-originated muons. 
The GF-driven high-brilliance beams of polarised positrons and muons may, therefore, help to reactivate R\&D programmes on: 
(1) a muon collider, see e.g.~\cite{Zimmermann:2018wfu},  
(2) a polarised lepton--hadron collider,
(3) fixed-target Deep Inelastic Scattering (DIS) experiments
and (4) a neutrino factory.

\subsection{High-purity neutrino beams}

The low-emittance muon beams can be used to produce high-purity neutrino beams. 
If  the muons are polarised, one can easily separate $\nu_{\mu}$ ($\bar{\nu}_{\mu}$) beams 
from the $\bar{\nu}_{e}$ ($\nu_{e}$) admixture on the bases of their respective angular distributions
resulting from the $(V-A)$-structure of the weak currents. 
Moreover, the neutrino and antineutrino bunches of each flavour can be separated with the $100\%$ efficiency based their timing. 
The fluxes of the neutrino and antineutrino beams are equal and they can be predicted to a per-mille  accuracy. 
The high-purity neutrino and antineutrino beams can be used for precision neutrino-physics measurements, e.g.\
of the CP-violating phase in the neutrino-mixing (PMNS) matrix.  

\subsection{Neutron and radioactive-ion beams}

The energy of the GF $\gamma$-rays can be tuned to excite the Giant Dipole Resonance (GDR) 
or fission resonances of large-$A$ nuclei, providing abundant sources of: 
(1) neutrons -- with the intensity up to $10^{15}$ neutrons per second (first-generation neutrons),
(2) radioactive and neutron-rich ions -- with the intensity up to $10^{14}$ ions per second. 
The above fluxes would  approach those of other European projects under construction, e.g.\ ESS, FAIR and EURISOL. 
The advantage of the GF sources is their high efficiency -- almost $10\%$ of the LHC RF power can  be  
converted into the power of the neutron and radioactive-ion beams.

\section{Project milestones}

On the path towards proving the feasibility of the GF concepts the following milestones have been set up:
\begin{enumerate}
 \item
 Demonstration  of efficient production, acceleration and storage of atomic beams 
 in the CERN accelerator complex. 
\item 
Development of the requisite GF simulation tools. 
\item
Successful execution of a GF Proof-of-Principle (PoP) experiment.    
\item 
Realistic assessment of the performance parameters of the GF research tools.  
\item 
Building up the physics cases for the research  programme and attracting wide scientific communities to 
use the GF tools in their respective research. 
\item
Elaboration of the GF Technical Design Report (TDR). 
\end{enumerate}

The above multi-step  path  to the feasibility proof of the GF concepts is planned such as
to minimise the infrastructure and hardware investments as well as  interference with the ongoing 
CERN research programme. Most of the PSI-beam tests are to be performed at the SPS and 
extrapolated to the LHC running conditions.
On the other hand, the ongoing CERN research programme may profit from the GF R\&D studies,
e.g.\  a successful demonstration fo the Doppler beam-cooling at the SPS might influence
the future LHC or AWAKE running scenarios. 
    
 The Gamma Factory R\&D programme started in 2017 and has already achieved the milestones no.~1. 
 Our current effort concentrates on the milestones no.~2 and 3.  

\section{Tests of production, acceleration and storage of PSI beams}

\subsection{SPS beam tests in 2017 and 2018}

In 2017 the ${\rm Xe}^{39+}$ beam was accelerated, stored in the SPS and studied at different flat-top energies \cite{CERN_courier:2017}. 
It turned out that the lifetime of this PSI beam was limited by electron stripping in collisions with residual gas in the beam pipe. 
The analysis of the measured lifetime was used to estimate the expected lifetimes of the ${\rm Pb}^{81+}$ and  ${\rm Pb}^{80+}$ beams in the SPS ring.
They turned out to be at least $100$ seconds, to be compared with only $40$ seconds needed to fill the SPS and to accelerate
the bunches up to the LHC injection energy.  This opened the possibility of injecting such beams to the LHC ring.    

In June 2018, both ${\rm Pb}^{81+}$ and ${\rm Pb}^{80+}$ beams were successfully 
injected to the SPS and accelerated to the proton-equivalent energy of $270\,$GeV \cite{CERN_news:2018}.  
The observed lifetimes of the ${\rm Pb}^{80+}$ and ${\rm Pb}^{81+}$ beams were $350\pm 50$ and $600 \pm 30$ seconds, respectively. 
The achieved intensities of these beams were in a good agreement with our expectations based on the calculations 
of the stripping efficiency for the initial ${\rm Pb}^{54+}$ beam. 
Finally, the achieved bunch intensity of the ${\rm Pb}^{81+}$ beam amounting to $8\times 10^9$ unit electric charges turned out to be 
comfortably higher than what was required for monitoring such bunches in both the SPS and LHC.  

\subsection{LHC beam tests with ${\rm Pb}^{81+}$ in 2018} 

On the 25\textsuperscript{th}  of July 2018, the  ${\rm Pb}^{81+}$ beam was injected for the first time to the LHC ring and accelerated
to the proton-equivalent energy of $6.5\,$TeV. The observed beam lifetime was $\sim 40$ hours.
The bunch intensity for the stable beam was 7 $\times$ 10$^9$ unit charges with $6$ bunches
circulating in the LHC. 
The date will be remembered for circulating, for the first time, a beam of ``atoms'' in the LHC \cite{CERN_news:2018}. 

\subsection{Lessons from SPS and LHC beam tests} 

The main outcome of the 2017 and 2018  GF test runs 
was the proof that the PSI beams can be formed, accelerated and 
stored in the existing LHC accelerator complex. Most of the operation aspects for such beams have been successfully tested. 
A very important achievement was the experimental demonstration that the bunches consisting of $10^8$ hydrogen-like
lead ions can be efficiently produced and maintained at the top LHC
energy with the lifetime and intensity reaching the GF requirements. 

These tests also validated our initial software tools, which will be 
further developed to extrapolate the production efficiencies of ${\rm Pb}^{81+}$ and  ${\rm Pb}^{80+}$ to 
arbitrary species of PSI and to predict their lifetimes in the SPS and the LHC. 

\section{Development of software tools}

The GF project requires new software tools to prepare the beam tests, to generalise 
their results to other beam species, to evaluate the GF intensity reach, to optimise 
the Interaction  Point (IP) of the laser photons and the PSI bunches, to study the internal dynamics of these 
bunches exposed to collisions with the laser light, to optimise the Doppler beam-cooling methods as well as to 
study the extraction and diagnostics of the GF primary and secondary beams. 

The existing software tools for simulations of electron-stripping in metallic foils and beam--gas collisions of highly charged ions \cite{Springer_2018} 
have been calibrated for their high-energy applications using striping efficiencies and beam lifetimes measured at the SPS and LHC
in the 2017 and 2018 tests, described in the previous section.

Two parallel projects are pursued to develop the requisite software tools  
for simulation studies of internal atomic beam dynamics: (1) based on a semi-analytical
approach and (2) based on Monte Carlo methods to study dynamics of individual ions. 
Their goal is to simulate the time evolution of the PSI-beam bunch parameters, 
such as the bunch emittances, the energy spread and the bunch length,
exposed to collisions with the laser photon pulses, as well as to study various beam-cooling scenarios. 

Two independent Monte Carlo generators for simulations of the photon-beam production in collisions of the PSI bunches with the laser pulses 
are being developed within the GF group \cite{Curatolo:2018pza,Curatolo:2018pos}. 
A major challenge of these projects is to incorporate, for the first time, resonant atomic-physics processes
into the frameworks developed by the particle physics community.  

Finally, new tools for simulations of the secondary-beams production need to be developed
or the existing ones ought to be upgraded appropriately. The recent progress in this domain includes
the ongoing  work  of the CERN SFT group on construction of a Monte Carlo 
generator for conversions of polarised photons into muon-pairs close to the production threshold. 

\section{SPS proof-of-principle experiment}

The most important and challenging  short-term  GF milestone is 
to validate the GF production scheme of the photon beams in 
a proof-of-principle (PoP) experiment in the SPS.
Its principal technical goals are:
\begin{itemize}
\item 
demonstration of a successful integration of the laser plus Fabry--P\'erot-cavity system into the storage ring of the high-energy hadron beam, 
including the interaction point  design and the radiation hardness of the laser system and its electronics;
\item 
optimisation of the laser-light frequency bandwidth to a given width of  the atomic excitation  and to the PSI-beam momentum spread;
\item
development of a collision scheme which maximises the rate of atomic excitations, 
including the optimal matching of the geometric characteristics of the PSI bunches to those of the laser pulses; 
\item
demonstration of the operational tools for the optimisation of the resonance excitation and its stability in time over many accelerator cycles;
\item 
demonstration of timing synchronisation of the laser pulses and the PSI bunches; 
\item 
study of extraction schemes of the $\gamma$-ray beam from the interaction-point  region and its collimation;
\item
development of  diagnostic methods of the PSI beams and the produced photon beams;
\item
tests of the Doppler beam-cooling schemes and evaluation of their reach in reducing the beam emittance.
\end{itemize}

For the PoP experiment, the PSI beam of the lithium-like lead, ${\rm Pb}^{79+}$, has been chosen. 
The photon beam is to be generated by using the atomic transition:
$1s^2 2s \;{}^2S_{1/2} \rightarrow1s^2 2p\;{}^2P_{1/2} $
with the energy difference of $230.76\,$eV, which can be excited with a $1030\,$nm pulsed laser 
by tuning the Lorentz $\gamma_L$-factor of the ${\rm Pb}^{79+}$-beam to the value of $\gamma_L \simeq 96$.
Such a  choice of the ion type and the atomic transition is constrained by the present quality 
of the SPS vacuum system.  The expected lifetime of such a beam in the SPS of $\sim 100$ seconds
is comfortably longer than the estimated beam-cooling time of $\sim 20$  seconds. 
The corresponding lifetime of the atomic excited state of ${\rm Pb}^{79+}$ is $76\,$ps.
This is by orders of magnitude longer than those for the radial atomic excitation which 
will be used to generate the target GF $\gamma$-ray beams.  
As a consequence, the PoP experiment at the SPS will test a more difficult GF configuration than the one planned for the LHC.  

\section{Summary}

The main goal of the GF initiative is to provide a variety of novel research tools at CERN. 
These tools can lead to diversification of the CERN scientific programme by  
opening new research opportunities in a broad domain of fundamental and applied physics. 
Such a diversification is becoming important in the present phase of high-energy physics research.
The GF project would make the full use of the existing CERN accelerator 
infrastructure  and would fit well to the CERN size and mission.
It would require relatively modest infrastructure investments. 
It can also potentially fit the time-gap before the necessary new acceleration technology 
becomes available to re-address the high-energy frontier of particle physics at a 
reasonable cost.  

The GF initiative requires a dedicated R\&D programme for an experimental proof of its concepts
as well as for a precise evaluation of intensity and quality aspects of its primary and secondary beams. 
The R\&D programme started in 2017 by creating the GF Working Group within the Physics Beyond Colliders Study Group at CERN.
It is organised such that its impact on the ongoing CERN research programme 
is minimal. It has already demonstrated the capacity of the CERN accelerators to produce,
accelerate and store the PSI beams. Now it has entered its second phase:  
preparation of the letter-of-intent for the GF proof-of-principle experiment 
in  the SPS and continuation of developing dedicated software tools for the project.  
The GF group has grown from a single-person initiative to a strong collaboration
involving  $62$ physicists and engineers from $23$ institutes in $10$ countries. 

\providecommand{\href}[2]{#2}
\end{document}